\documentclass[useAMS]{mn2e}
\usepackage{natbib_jrm,graphicx}
\bibpunct[, ]{(}{)}{;}{a}{}{,}
\def \aj {AJ}
\def \mnras {MNRAS}
\def \pasp {PASP}
\def \apj {ApJ}

\def \apjl {ApJL}
\def \aap {A\&A}
\def \nat {Nature}
\def \araa {ARAA}
\def \iaucirc {IAUC}

\def\lesssim{\mathrel{\hbox{\rlap{\hbox{\lower4pt\hbox{$\sim$}}}\hbox{$<$}}}}
\def\gtrsim{\mathrel{\hbox{\rlap{\hbox{\lower4pt\hbox{$\sim$}}}\hbox{$>$}}}}
\begin{document}
\title{Hubble Space Telescope imaging of the progenitor sites of six nearby 
core-collapse supernovae}

\author[Maund and Smartt]
{J.R. ~Maund$^{1}$\thanks{Email: jrm@ast.cam.ac.uk} and S.J. ~Smartt$^{2}$\\
$^{1}$Institute of Astronomy, University of Cambridge, Madingley Road, Cambridge 
CB3 0HA, England, U.K.\\
$^{2}$Department of Physics and Astronomy, Queen's University Belfast, Belfast, 
BT7 1NN, Northern Ireland, U.K.}

\maketitle
\begin{abstract}
The search for the progenitors of six core-collapse supernovae (CCSNe) in archival HST WFPC2 pre-explosion imaging is presented here.  These SNe are 1999an, 1999br, 1999ev, 2000ds, 2000ew and 2001B.  Post-explosion imaging of the SNe, with the HST ACS/WFC, has been utilised with the technique of differential astrometry to identify the progenitor locations on the pre-explosion imaging.  SNe 1999br, 1999ev, 2000ew and 2001B are recovered in late time imaging and an estimate of the progenitor location on the pre-explosion imaging, with sub-pixel accuracy, has been made.  Only the progenitor of the type II-P SN 1999ev has been recovered, on pre-explosion F555W imaging, at a $4.8\sigma$ significance level.  Assuming a red supergiant progenitor, the pre-explosion observation is consistent with $M_{ZAMS}=15-18M_{\odot}$.  The progenitors of the other five SNe were below the $3\sigma$ detection threshold of the pre-explosion observations.  The detection thresholds were translated to mass limits for the progenitors by comparison with stellar evolution models.  Pre-explosion observations of the peculiarly faint SN 1999br limit the mass of a red supergiant progenitor to $M_{ZAMS} < 12M_{\odot}$.  Analysis has been extended, from previous studies, to include possible detections of high-$\mathrm{T_{eff}}$, high-mass stars by conducting synthetic photometry of model Wolf-Rayet star spectra.  The mass limits for the type II-P SNe 1999an and 1999br are consistent with the previously determined mass limits for this type of SN.  The detection limits for the progenitors of the type Ibc SNe (2000ds, 2000ew and 2001B) do not permit differentiation between high-mass WR progenitors or low mass progenitors in binaries.
\end{abstract}

\section{Introduction}
Supernovae (SNe) are associated with the end points of the evolution
of particular types of stars.  Stars with initial masses $>8M_{\odot}$ are
believed to undergo a core-collapse induced explosion at the end of their
lives \citep{heg03}.  SNe are important drivers of the chemical and
physical evolution of their host galaxies: distributing heavy elements
newly synthesised in the explosion and injecting kinetic and radiative
energy into the surrounding medium \citep{thi03}.  The type of the SN depends
on the evolutionary status of the progenitor object just prior to
core-collapse.  The observational identification of these objects, and
determination of their nature, is of importance to the study of the
evolution of massive stars and the evolution of SNe \citep{hampi02}.\\
SNe are principally classified by the absence (Type I) or presence (Type II)
of {\sc Hi} features in their early time optical spectral.  SNe are
further sub-classified by other spectral features or the shapes of their
light curves.
The type Ia SN class is very homogeneous, characterized by
the presence of a Si II absorption feature \citep{fili97}.  SN Ia occur in
old-intermediate age stellar populations (in E, S0 and the bulges of S
galaxies) composed of low-intermediate mass stars.  SN Ia are
thought to arise from the thermonuclear explosion of either C-O white
dwarfs accreting material from a binary companion or the coalescence of
two C-O white dwarfs in a close binary \citep{branIa}.
The Si II absorption feature is not seen in types Ib/c.  SN Ib show
spectral features due to He, while SN Ic are He deficient.  These SNe,
along with all type II SNe, are seen in young blue star forming regions in
the spiral arms of late type galaxies and are associated with massive star
progenitors ($M_{ZAMS} > 8M_{\sun}$) undergoing a core-collapse
induced explosion.  The absence of H, and the subsequent absence of He in
SN Ic, is associated with increasing mass loss, and hence initial mass, of
the progenitor or enhanced mass loss due to a binary interaction
\citep{podsiIbc04}.
Type IIP/L SNe show prominent features due to {\sc H}.  SN IIP exhibit a
plateau feature in their light curves, due to the expansion and
recombination of a large {\sc H}-rich envelope, as well as H Balmer P-Cygni
profiles in their optical spectra \citep{hamobs03}.  The progenitors for these
objects are likely to be stars towards the lower range of the masses that will 
produce a core-collapse SN, for which mass loss is not
significant allowing the H-envelope to be retained.
\citet{smartt01du} presented initial mass limits,
from pre-explosion imaging, for a number of type IIP SN progenitors,
suggesting the progenitors were all stars with initial mass
$<15M_{\sun}$.  The detected progenitor of the IIP SN2003gd an M0
supergiant, had an initial
mass $8^{+4}_{-2}M_{\sun}$, consistent with the theoretical minimum
initial mass for a core-collapse SN \citep{smartt03gd}.
SN IIL are thought to arise for slightly more massive stars
($\sim20M_{\sun}$) for which mass loss is much more important.  In these
cases the stars have retained an H-envelope of mass $\sim1-2M_{\sun}$,
which is not massive enough to produce a plateau and so the SN light curve
decays linearly \citep{turclass03}.  In addition there are a number of type II subclassifications which are classified by 
spectroscopic features.
Type IIb SNe transform from a type II to a Ib SN, of which SN1993J was the
class prototype. The transition is due the retention of thin veil of the H
envelope which gives rise to the initial
classification of type II, but upon dissipation transforms to Ib.
 Low velocity IIn SNe exhibit narrow
spectral lines.  This may be due to collision of the SN with a dense
circumstellar medium (e.g. 1998S, \citealt{lentz98s}) or the eruptions of Luminous Blue 
Variables (e.g. 1997bs, \citealt{vandyk97bs}) which appear similar to CCSNe.\\
Core-collapse SNe (CCSNe) are a heterogenous group due to the different
permutations of the explosion.  In the cases of single star progenitors the
sequence IIP$\rightarrow$IIL$\rightarrow$IIb$\rightarrow$Ib/c \citep{heg03}
can be considered to be due to increasing initial mass, and hence mass
loss during the final stages of evolution.  The outer H rich and
subsequently He rich layers are stripped due to the increasing levels of
mass loss.  The role of binaries, and in particular mass and angular
momentum exchange,  provide alternative routes for different mass
progenitors to appear as the same SN type \citep{maund93j}.
The direct determination of the nature of an SN progenitor requires
observations of the progenitor prior to explosion.  In most cases this is from 
fortuitous imaging available from telescope archives which was acquired
for some other purpose, but has also imaged the SN location.  Given the
rarity and randomness of CCSNe it is necessary to look to nearby
galaxies ($\mathrm{<20Mpc}$) in order to observe enough events at a
feasible rate, coupled with a reasonable probability of detecting the progenitor
and resolving it from neighbouring stars.  A multi-wavelength deep
imaging survey of nearby galaxies has been undertaken by the authors to
improve the likelihood of appropriately deep and high resolution pre-explosion
data being available for future nearby SNe.\\
In only three cases have the 
progenitors of definite CCSNe, with their parameters fully
determined, been identified in pre-explosion frames:
1987A \citep{gil87a}, 1993J \citep{alder93j} and
2003gd \citep{smartt03gd}.  SN1987A in the Large Magellanic Cloud (LMC)
was noted as a peculiar type II SN and the progenitor star was identified
as Sk $-69\degr202$ \citep{walb87a} a B3Ia star.  SN1993J was the
prototypical type IIb SN, in M81, and the progenitor was identified as
K0Ia star with an unknown ultraviolet component \citep{alder93j}.
Neither of these progenitors is consistent with the canonical
M-supergiant progenitor predicted for these types of SNe.  In both cases,
however, there is evidence that binarity plays a particularly important
role. \citet{podsi87a} describe a binary model for the progenitor of SN
1987A and \citet{maund93j} report the discovery of a blue supergiant
companion to the progenitor of SN 1993J.  The detected progenitor of SN 
2003gd was the first red supergiant progenitor observed for a type IIP SN.
In a number of cases pre-explosion imaging of the SN location has been
available in a number of archives, in particular that of the Hubble Space
Telescope (HST), but the pre-explosion images were not deep enough to detect
the progenitor.  In these cases \citet{smartt99em,smartt99gi,smartt01du}
have placed brightness limits on the progenitors given the limiting
magnitude of the pre-explosion frames at the SN position.  The
brightness limits have then  been translated to mass limits by plotting
the detection threshold on the Hetrzsprung-Russell diagram and determining
the mass range for which the progenitor would not have been detected
in the pre-explosion image.  \citet{smartt99em,smartt99gi,smartt01du} have 
used post-explosion imaging of the SN, with the technique of differential 
astrometry, to locate the SN position on pre-explosion images to sub-pixel 
accuracy.  \citet{vandykprog} have similarly attempted to identify progenitors 
in pre-explosion imaging.  They have used the reported positions of SNe, such 
as given in IAU Circulars.  The correct identification of progenitors is 
limited, however, by the poor pointing accuracy of HST (requiring the 
astrometric re-calibration of pre-explosion frames) and uncertainties in 
the reported positions of SNe.\\
The parameters of the progenitor may also be estimated, using various
calibrations, from the evolution of the SN itself.  The light
curves and velocity evolution of the SNe have been used, with particular
theoretical calibrations, to determine the ejected and Ni masses
\citep{hamobs03}.  Epochs of mass loss have been probed by observing
the variations in the radio light curves of certain SN due to the interaction
of the SN with the progenitor's circumstellar medium\citep{ryder01ig}.
Direct detection of the progenitor, whilst being a reliable method of
determining the progenitor parameters, is also a useful check and
calibration for the other techniques for determining the properties of the progenitors
from the SN.\\
In this paper we present the search for the progenitors of six nearby CCSNe: 3
type IIP SNe (1999an, 1999br and 1999ev), 2 type Ib SNe (2000ds and 2001B)
and 1 type Ic (2000ew).  Fortuitous pre-explosion WFPC2 imaging for these
SNe is available in the Hubble Space Telescope (HST) archive.  The host
galaxies of these SNe lie within 26Mpc, increasing the likelihood of detecting 
and resolving individual stars.  The properties of these SNe, and
their host galaxies\footnote{http://www-obs.univ-lyon1.fr/hypercat/}, are presented in table \ref{Snhosts}.
\begin{centering}
\begin{table}
\label{Snhosts}
\caption{A table listing the properties of the SNe and their host
galaxies discussed in this study.}
\begin{tabular}{llcclr}
\hline\hline
SN & SN type & Host galaxy & Type & $\mu$ & $i$\\
\hline
1999an  &II    & IC 755  &  SBb     & 31.79  &$90$\\
1999br  &IIpec &NGC 4900 &  SBc     & 30.82  &$16.5$\\
1999ev  &II    &NGC 4274 &  SBab(r) & 30.96  &$71.8$\\
2000ds  &Ib    &NGC 2768 &  E-Sa    & 31.84  &$90$\\
2000ew  &Ic    &NGC 3810 &  Sc      & 30.90  &$47.8$\\
2001B   &Ib    & IC 391  &  Sc      & 32.07  &$11.14$\\
\hline\hline
\end{tabular}
{\em $\mu$} Kinematical distance modulus quoted by LEDA, assuming 
$\mathrm{H_{0}=70kms^{-1}}$.\\
{\em $i$} Inclination angle of the plane of the galaxy quoted by LEDA.
\end{table}
\end{centering}
The analysis of pre- and post-explosion imaging of these SNe (photometry,
differential astrometry and determination of the limiting magnitudes)
is presented in \S2.  \S3 presents the results of the
observations for each of the six SNe in this study.  These results will be
discussed, in the context of the parameters of the progenitors, in
\S4.  We adopt the technique of \citet{smartt99gi, smartt99em,smartt01du}
to place mass limits on the progenitors.  \citet{vandykprog} have attempted 
progenitor detection of these SNe in the publicly available WFPC2 images.  
They claim possible detections of the progenitors of SNe 1999br, 1999ev and 
2001B.  We show, however, that the only progenitor recovered is that of 
the type II SN 1999ev.
\section{Observations and Data Analysis}
\subsection{Pre- and post-explosion WFPC2 images}
\label{anawfp2}
Hubble Space Telescope (HST) Wide Field Planetary Camera 2 (WFPC2)
observations of the SN location acquired prior to explosion were
identified using the ST-ECF {\sc Astrovirtel} {\it Querator} tool\footnote{http://archive.eso.org/querator/}.  This tool identifies
exposures for which the target falls on the field of view,
taking into account the peculiar mosaic shape of the WFPC2.  The
data was passed through the On-the-fly-recalibration (OTFR) pipeline
and retrieved from the ST-ECF archive.  A journal of these observations
is presented as table \ref{preexptable}.  In addition a number of post-explosion
WFPC2 images of the SNe were available and these are listed in table \ref{postexptab}.
The post-explosion frames were reduced and analysed in a similar manner
to the pre-explosion images.  Individual cosmic ray exposures were
combined under {\sc Iraf} using the {\sc Stsdas} {\it crrej} routine.
Image shifts, in the cases of dithered observations, were adopted from
the WFPC2-B association tables and applied to bring the images to a
common origin.  Shifts were made with the {\sc Iraf} task {\it imshift}.
The combined and distortion corrected images were photometered using
the {\sc Iraf} implementation of the {\sc DAOphot} package \citep{stet87}.
Point spread function (PSF) models, taking into account variations
across and between individual chips, were created using the {\sc TinyTim}
program \citep{tinytim04} and were used in the photometry task ALLSTAR.
Aperture and charge transfer efficiency corrections
\citep{holsper95,dolp00cte} were applied to the output photometry.
The output STMAG magnitudes were converted to the VegaMag system using
the latest zeropoints of \citet{dolp00cte}\footnote{http://www.noao.edu/staff/dolphin/wfpc2\_calib}.  The positions
of stars, output by {\sc DAOphot}, were utilised to recalibrate
the absolute astrometry of the WFPC2 frames and align groups of images of the 
same galaxy.  The recalibration of the absolute astrometry and the technique 
of differential astrometry is discussed in \S\ref{anaast}.
\begin{centering}
\begin{table*}
\caption{\label{preexptable}Summary of pre-explosion data}
\begin{tabular}{clllrl}
\hline\hline
SN     & Date of Observation & Dataset    & Filter & Total Exposure & HST program\\
       &                     &            &        & Time (s)       &
  \\
\hline
1999an & 1995 Jan 05         &  U29R3R01/02 & F606W  & 160            & GO-5446
  \\
\\
1999br & 1995 Jan 29         &  U29R2N01/02 & F606W  & 160            & GO-5446
  \\
\\
1999ev & 1995 Feb 05         &  U2JF0101/02/03 & F555W  & 296            & GO-5741
  \\
\\
2000ds & 1995 Dec 04         &  U2TV1803 & F555W  & 160            & GO-5999
  \\
       & 1995 Dec 04         &  U2TV1801/02 & F814W  & 320            & GO-5999
  \\
       & 1998 May 20         &  U3M71605/06 & F555W  &1000            & GO-6587
  \\
       & 1998 May 20         &  U3M71608/09 & F814W  &2000            & GO-6587
\\
\\
2000ew & 1994 Nov 04         &  U29R1R01/02 & F606W  & 160            & GO-5446
  \\
\\
2001B  & 1994 Feb 21         &  U2920B01 & F555W  &  70            & GO-5104
  \\
\hline\hline
\end{tabular}
\end{table*}
\end{centering}
\subsection{Post-explosion ACS images}
\label{anaacs}
Post-explosion observations of these SNe were conducted as part of our
program GO9353 (PI: S. Smartt)
for HST cycle 11, with the HST Advanced Camera for Surveys (ACS) Wide
Field Camera (WFC).  Observations were conducted in three broad-band
filters: F435W, F555W and F814W.  A journal of these observations is
presented in table \ref{postexptab}.  Images, passed through the OTFR pipeline,
were retrieved from the ST-ECF archive. A grid of PSF models for both WFC
chips and for each set of filters used  were constructed using the {\sc
TinyTim} program.  Photometry was then conducted using the {\sc DAOphot}
{\sc ALLSTAR} routine.  Empirical aperture corrections were calculated,
 for each frame individually, to a radius of 0.5\arcsec.  A correction was
made for the charge transfer efficiency following the prescription of
\citet{riesscte}, adopting the updated distortion coefficients\footnote{http://www.stsci.edu/hst/acs/analysis/PAMS} of \citet{coxdistort}.
The stellar photometry was then converted to standard Johnson-Cousins
{\it B},{\it V} and {\it I} magnitudes utilising the relationships of
\citet{holsphot95}, since the ACS photometric system is calibrated to
within 0.03 magnitudes of the corresponding filters in the WFPC2
photometric system \citep{acshand}.
\begin{centering}
\begin{table*}
\caption{\label{postexptab}Summary of post-explosion data}
\begin{tabular}{llrrrrr}
\hline\hline
SN      & Date of Observation & Dataset    & Filter & Total Exposure & Instrument & HST      \\
        &                     &            &        & Time (s)       &
  &program   \\
\hline
1999an  & 2002 Jul 12         &  J8DT05030 & F435W  &  400           & ACS/WFC
  & GO-9353  \\
       & 2002 Jul 12         &  J8DT05010 & F555W  &  430           & ACS/WFC
  & GO-9353  \\
        & 2002 Jul 12         &  J8DT05030 & F814W  &  430           & ACS/WFC
  & GO-9353  \\
\\
1999br  & 2002 Jun 20         &  U6EA8601/02 & F450W  &  460           & WFPC2
  & GO-9042  \\
        & 2002 Jun 20         &  U6EA8603/04 & F814W  &  460           & WFPC2
  & GO-9042  \\
        & 2002 Dec 23         &  J8DT06030 & F435W  &  400           & ACS/WFC
  & GO-9353  \\
        & 2002 Dec 23         &  J8DT06010 & F555W  &  430           & ACS/WFC
  & GO-9353  \\
        & 2002 Dec 23         &  J8DT06020 & F814W  &  430           & ACS/WFC
  & GO-9353  \\
\\
1999ev  & 2002 Dec 31         &  J8DT03030 & F435W  &  400           & ACS/WFC
  & GO-9353  \\
        & 2002 Dec 31         &  J8DT03010 & F555W  &  450           & ACS/WFC
  & GO-9353  \\
        & 2002 Dec 31         &  J8DT03020 & F814W  &  450           & ACS/WFC
  & GO-9353  \\
\\
2000ds  & 2002 May 31         &  J8DT02030 & F435W  &  680           & ACS/WFC
  & GO-9353  \\
        & 2002 May 31         &  J8DT02010 & F555W  &  700           & ACS/WFC
  & GO-9353  \\
        & 2002 May 31         &  J8DT02020 & F814W  &  700           & ACS/WFC
  & GO-9353  \\
\\
2000ew  & 2001 Nov 07         &  U6EA5401/02 & F450W  &  460           & WFPC2
  & GO-9042  \\
        & 2001 Nov 07         &  U6EA5403/04 & F814W  &  460           & WFPC2
  & GO-9042  \\
        & 2002 Jun 26         &  J8DT04030 & F435W  &  400           & ACS/WFC
  & GO-9353  \\
        & 2002 Jun 26         &  J8DT04010 & F555W  &  430           & ACS/WFC
  & GO-9353  \\
        & 2002 Jun 26         &  J8DT04020 & F814W  &  430           & ACS/WFC
  & GO-9353  \\
\\
2001B   & 2002 Jun 09         &  J8DT01020 & F435W  &  800           & ACS/WFC
  & GO-9353  \\
        & 2002 Jun 09         &  J8DT01010 & F555W  &  700           & ACS/WFC
   & GO-9353  \\
        & 2002 Jun 09         &  J8DT01030 & F814W  &  700           & ACS/WFC
  & GO-9353  \\
\hline\hline
\end{tabular}
\end{table*}
\end{centering}
\subsection{Astrometry}
\label{anaast}
The pointing accuracy of HST, and hence the absolute astrometric accuracy
of HST imaging, is of the order 1-2\arcsec \citep{smartt01du}.  In
addition the accuracy of the reported SN positions is $\sim0.5-1\arcsec$.
This amounts to a potential error of 3\arcsec or 30 WF pixels on WFPC2.
\citet{vandykprog} have utilised catalogued positions of stars which
fall on the pre-explosion images to recalibrate the absolute astrometry
and compensate for the uncertainty in HST pointing.  The uncertainty of
the reported SN position is, however, still present.  In crowded fields,
therefore, better astrometry is necessary to provide a more secure
identification of the progenitor object.  \citet{smartt99gi,smartt99em,
smartt01du} have utilised purposefully acquired post-explosion imaging to 
provide an accurate position for the progenitor object on pre-explosion images
using the technique of differential astrometry.  The transformation
between the pre- and post-explosion frames, using stars common to both
images, allows the determination of the SN location on the pre-explosion image.  
In this study the technique of differential astrometry
is preferentially used if the SN has been recovered in the
post-explosion frames.  The astrometry of the pre-explosion WFPC2 images
were re-calibrated, in cases when the SN was not recovered, using stars
catalogued in the USNO-B1.0 \citep{monet03} and Automated Plate Machine
(APM, \citealt{apm92}) catalogues.  The astrometric precisions of these
catalogues are quoted as 0.2\arcsec and 0.1\arcsec respectively.
The pre- and post-explosion images were aligned, by identifying common
stars in both frames, using the positions calculated by {\sc DAOphot}.
The transformations between images were calculated using the {\it geomap}
task in {\sc Iraf}.  The absolute astrometry of the images
were
recalibrated using the {\it ccmap} and {\it ccsetwcs} tasks.  All
positions quoted in this study are J2000.0.  The plate scales of the WF and 
PC chips of WFPC2 are $0.1\arcsec$ and $0.05\arcsec$ respectively.  The ACS 
WFC has the same plate scale as the WFPC2 PC chip.  Photometry and astrometry 
were conducted on the original data frames, but for the purposes of comparison 
all the images presented here have been translated, rotated and scaled to be 
consistent with the coordinates of the ACS WFC post-explosion images.   

\subsection{Detection limits of pre-explosion WFPC2 images}
\label{seclim}
In the cases when the progenitor was not detected in the pre-explosion
frames the detection limit of the images was determined in order to place
a brightness limit on the progenitor.  The brightness of the background was 
measured using aperture photometry.  The properties of the background were 
then used to estimate the stellar flux required for a detection of 3$\sigma$.
The signal-to-noise ratio for a photometered star is given by:

\begin{equation}
S/N=\frac{F_{star}}{\sqrt{F_{star}+Q}}
\end{equation}

where $F_{star}$ is the flux of the star in electrons and Q represents
the noise contribution from the background and readnoise.
Q is given by:

\begin{equation}
Q=A_{star}\left(1+\frac{A_{star}}{A_{sky}}\right)(F_{sky}+R^{2})
\end{equation}

$F_{sky}$ is the average flux of a sky pixel.  $A_{star}$ and $A_{sky}$
are the respective areas of the aperture and sky annulus in pixels.  $R$
is the read noise in electrons, which for the WFPC2 images summed with {\it crrej}
is given by the sum, in quadrature, of the read noises of the individual frames.  
The calculation for the signal-to-noise ratio may be inverted, given the
sky background as measured by the aperture photometry, to determine the
flux of star at the detection threshold.

\begin{equation}
F_{star}=\frac{1}{2}S/N\sqrt{S/N^{2}+4Q}+\frac{1}{2}S/N^{2}
\end{equation}

In this study the threshold will be taken as $S/N=3$.
The limiting magnitude was then calculated in the standard manner taking into
account the corrections for the VegaMag zeropoint, aperture size and charge transfer efficiency.

\subsection{Estimates of reddening}
\label{secred}
The post-explosion photometry of stars, in the direct vicinity of the SN, was 
utilised to determine a reddening along the line of sight.  This 
simultaneously takes into account the contribution of the internal reddening and the foreground reddening.  Independent values of the 
fore-ground reddening were adopted from the NASA/IPAC Extragalactic Database 
(NED)\footnote{http://nedwww.ipac.caltech.edu/}, after \citet{schleg98}.  
The apparent colours, $B-V$ and $V-I$, of stars within $6\arcsec$ of each SN 
were compared against a standard supergiant intrinsic colour sequence, 
from \citet{drill00}.  The displacement of the stars' apparent colours from 
the intrinsic supergiant colour sequence was calculated along the reddening 
vector, adopting a standard \citet{ccm89} and \citet{odonn94} reddening law.  
Intermediate values in the supergiant intrinsic colour sequence were 
estimated using linear interpolation.   The displacement $D$ between the apparent 
and intrinsic colours of the stars was related to the reddening by:
\begin{eqnarray}
D^{2} & = &\left( \left( B-V\right)-\left(B-V\right)_{0}\right)^{2}+\left( \left
( V-I\right)-\left(V-I\right)_{0}\right)^{2}\nonumber\\
      & = & \left(E\left(B-V\right)\right)^{2}+\left(E\left(V-I\right)\right)^{2
}\nonumber\\
      & = & \left(1+\alpha\right)^{2}\left(E\left(B-V\right)\right)^{2}
\end{eqnarray}
For a \citet{ccm89} reddening law, with $R_{V}=3.1$, $\alpha=0.62$.  A 
weighted mean value of the reddenings of the all the stars was adopted as 
the total reddening towards the SN.  The extinctions in the 
WFPC2 band-passes were calculated, from $A_{V}$, using the ratios of extinctions of \citet{schleg98}.  
The precision of this technique is limited by the quality of the photometry 
and hence the 
number of bright nearby stars considered.  This technique indicates, 
however, when the reddening appropriate for the SN is much higher than the 
foreground reddening alone.  This is particularly important when considering the 
complicated line of sight to SNe in galaxies with large inclinations.
\section{Observational Results}
\label{obssec}
%
\subsection{SN 1999an}
\label{99anobssec}
SN 1999an was discovered on 1999 March 7.83 UT \citep{99aniauc1} in the SBb
galaxy IC 755 and the presence of strong H Balmer P Cygni profiles showed
it to be a classical type II SN.  \citet{99aniauc2} noted the position of
the SN as $\mathrm{\alpha=12^{h}01^{m}10^{s}.57 , \delta=+14\degr 06 \arcmin 12
\arcsec .3}$.  \citet{vandykprog} independently quote a position
$\mathrm{\alpha=12^{h}01^{m}10^{s}57 , \delta=+14\degr06\arcmin11\arcsec.1}$ with a
conservative uncertainty of $\pm0.4\arcsec$.
A 160s pre-explosion F606W WFPC2 image, acquired 3.2 years prior to discovery 
(program GO-5446, PI: G. Illingworth),
was available in the HST archive.  The SN location was re-imaged with the ACS/WFC, in three 
colours, 3.3 years after discovery.  The coordinate transformation between 
the pre-explosion frame and the post-explosion F555W image was calculated, 
with an uncertainty of 0.46 WF pixels (WFPC2) or $0.046\arcsec$.  The 
pre-explosion frame, due to a more advantageous orientation including six 
USNO-B1.0 stars, was astrometrically recalibrated.  The absolute astrometric 
accuracy was $0.42\arcsec$.  The SN location occurred on the WF2 chip of the 
pre-explosion image.  The total uncertainty of the SN position was 
$0.62\arcsec$ or 6 WF pixels.  The SN location on the pre- and post-explosion 
images is shown as figure \ref{99animages}.
\begin{figure*}
\caption{Pre- and post-explosion images of the site of SN 1999an in IC 755. 
(a) WFPC2 F606W 160s exposure, acquired on 1995 January 5, scaled and rotated 
to coordinates of ACS/WFC observations , (b) ACS/WFC F555W 430s exposure 
acquired on 2002 July 12 and (c) an ACS/WFC F814W 430s exposure acquired at 
the same epoch as the F555W image.  The uncertainty in the SN position, from the 
quoted absolute astrometry, is indicated by the black circle. {\bf - see 
jpeg}}
\label{99animages}
\end{figure*}
There is no noticeable difference between the pre- and post-explosion images 
within the error circle of the SN location.  This suggests that the 
progenitor is not detected in the pre-explosion image and the SN is not 
recovered on the post-explosion image to limiting magnitudes of $m_{F435W}=26.2\pm0.5$, $m_{F555W}=26.0\pm0.4$ and $m_{F814W}=25.5\pm0.4$.  The limiting magnitude at the SN 
location on the pre-explosion image was determined to be $m_{F606W}=24.7\pm0.2$.  IC 
755 is almost exactly edge on with evidence for bar feature \citep{lutt00}.  
This implies that the internal reddening towards SN1999an and its progenitor 
is likely to be higher than the NED quoted fore-ground reddening $E(B-V)=0.032$.  The reddening, 
including the internal contribution, determined from nearby stars was 
$E(B-V)=0.13\pm 0.06$.  This implies an absolute 
extinction of $A_{F606W}\approx A_{V}=0.42$.  At the distance of IC 755, 
22.8Mpc, this gives a limiting absolute magnitude of $M_{F606W} \ge -7.48$.
%
\subsection{SN 1999br}
\label{99brobssec}
\begin{figure}
\begin{center}
\rotatebox{-90}{
\includegraphics[width=5.8cm]{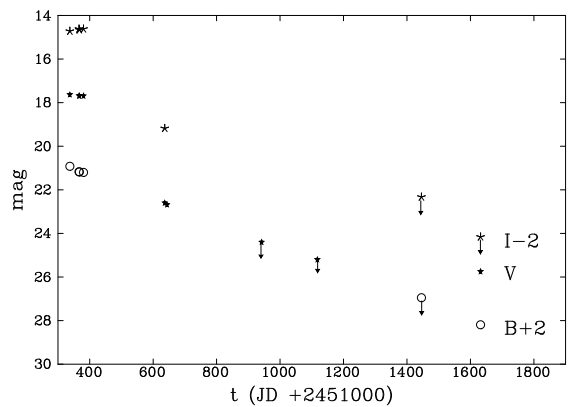}
}
\caption{A figure showing the light curve of SN 1999br.  Photometry of the 
early phases of the SN (JD $<$ 24552200) is from \citet{past99br}.  Late 
time magnitudes and magnitude limits are from HST WFPC2 (GO 9042) and ACS 
(GO 9353) observations.}
\label{lcfig}
\end{center}
\end{figure}

SN 1999br was discovered on 1999 April 12.4 in the galaxy of NGC 4900 by
\citet{99briauc1}.  The presence of broad Balmer lines showed this to be
an early Type II SN \citep{99briauc2a}, although the the relative strengths
of these lines indicated that it was sub-luminous \citep{99briauc2b}.
\citet{99briauc3} found that the expansion velocities of this SN were
much lower than normal Type II SNe and similar to the low-energy SN 1997D. 
Photometric observations by \citet{past99br} showed a plateau in 
the light curve, classifying 1999br as a type II-P SN.  \citet{99briauc1} gives the location of the SN as
$\mathrm{\alpha=13^{h}00^{m}41^{s}.80 , \delta=+2\degr29\arcmin 45\arcsec.8}$.
The SN position was imaged with HST WFPC2, with a 160s F606W exposure,
4.16 years prior to explosion (program GO-5446, PI: G. Illingworth).  WFPC2 and 
ACS images of the SN were acquired
3.19 and 3.70 years after explosion respectively.  The absolute astrometry 
of the three colour ACS imaging was recalibrated using APM 
catalogue stars, with a 
final uncertainty on the SN position as $\pm 0.7\arcsec$.
A V-band image of NGC 4900 from the TNG, acquired 101 days after discovery with the OIG instrument \citep{past99br}, was utilised
to provide a separate differential astrometric calibration.  The SN 
position was
located, using the TNG image, on the post-explosion ACS frames with a precision
of 6 ACS/WFC pixels (the combined error of the ``seeing'' $\sigma$ of the TNG image and the uncertainty in the transformation between the TNG and ACS images).  The SN is not significantly recovered on the 
post-explosion WFPC2 images with limiting magnitudes $m_{F450W}(3\sigma)
=24.95$ and $m_{F814W}(3\sigma)=24.33$.  An object is recovered in the post-explosion 
ACS F435W and F555W images within the error circle of the differential 
astrometry from the TNG image.  This object has magnitude $m_{F435W}=26.19\pm0.20$ ($4.3\sigma$)
and $m_{F555W}=25.70\pm0.18$.  This object does not appear significantly in the 
F814W image to a $3\sigma$ detection limit of $m_{F814W}=26.18$.   Figure 
\ref{lcfig} shows 
these magnitudes are consistent with the measured light curve of SN1999br of \citet{past99br}.   
The post-explosion ACS and WFPC2 images were 
aligned with the pre-explosion WFPC2 frame.  The SN location on the TNG,
ACS and both the pre- and post-explosion WFPC2 images is shown as figure
\ref{99brstamp}.  The SN location falls on WF3 chip on the pre-explosion 
imaging with a positional uncertainty of 3 WF pixels.  There is an 
indication of an object at the centre of the error circle, but it is just below the 
$3\sigma$ detection limit. 
\citet{vandykprog} suggest an object $0.7\arcsec$ from the SN position, as determined from the post-explosion TNG image, and detected at $3.3\sigma$ is the likely progenitor.  This object is indicated on figure \ref{99brstamp}b.  The post-explosion ACS/WFC observations show two extended objects either side of the transformed position of the centroid of this object on the pre-explosion frame.  This suggests that the object imaged on the pre-explosion frame is merely the unresolved combination of these two features and not the progenitor.
The limiting magnitude 
at the position of SN1999br on the pre-explosion image was determined to be $m_{F606W}=24.91$.  The reddening of nearby stars was determined to be $E(B-V)=0\pm0.05$.  NGC 4900 has an inclination of $16.5\degr$ and this 
reddening is consistent with a negligible contribution from internal 
extinction expected for such an almost face on spiral.  We adopt, therefore, the NED quoted fore-ground extinction of $E(B-V)=0.024$ towards NGC 4900.  The limiting absolute magnitude for SN 1999br was, therefore, $M_{F606W}\ge-5.98$.
\begin{figure*}
\caption{Pre- and post-explosion imaging of the site of SN 1999br in the NGC 4900.  
(a) Post-explosion ACS/WFC F555W image showing an object detected at the location 
of SN1999br. (b) Pre-explosion WFPC2 F606W image.  The SN location is indicated, 
with the positional uncertainty of $0.3\arcsec$ (6 ACS/WFC pixels or 3 WFPC2 WF3 pixels), by the circle.  The cross hairs indicate the object suggested by \citet{vandykprog} as the likely progenitor.  (c) TNG/OIG V-band image of SN 1999br.  
The position of the SN on this image was used to provide an independent location of the 
SN on pre-explosion and later post-explosion imaging. (d) Post-explosion WFPC2 PC F814W 
image of the SN location.  The SN is not detected, 3.14 years after discovery, with a 
limiting magnitude of $m_{F814W}(3\sigma)=24.33$. {\bf - see jpeg}}
\label{99brstamp}
\end{figure*}
%
\subsection{SN 1999ev}
\label{99evobssec}
SN 1999ev was discovered by T. Boles \citep{99eviauc1} on 1999 November 7.225.
The position of the SN was 
$\mathrm{\alpha=12^{h}19^{m}48^{s}.20 , \delta=+29\degr37\arcmin21\arcsec.7}$
\citep{vandykprog}, in the SBab galaxy of NGC 4274 (
$v_{vir}=1089\mathrm{kms^{-1}}$). \citet{99eviauc1a} noted a strong 
$\mathrm{H\alpha}$ P-Cygni profile consistent with a normal type II SN 
past maximum.  Photometric observations\footnote{http://www.rochesterastronomy.org/snimages/sn1999/sn1999ev.html} of this SN show a plateau in the light curve and, hence, imply a II-P classification.  The SN location had been imaged with a 296s F555W exposure 
on 1995 February 5 (program GO-5741, PI: J. Westphal).  Post-explosion ACS WFC imaging, acquired on 2002 
December 31, shows a new feature within the error circle of the SN position 
as determined from absolute astrometry.  The pre- and post-explosion images 
are shown as figure \ref{99evstamp}.
\begin{figure*}
\caption{Pre- and post-explosion imaging of the site SN 1999ev in NGC 
4274.  (a) Pre-explosion WFPC2 F555W image.  The progenitor was identified 
in this image, to a positional accuracy of 0.2 WF pixels on the 
pre-explosion image.  The progenitor had magnitude 
$m_{F555W}=24.64\pm0.17$, detected at $4.8\sigma$.  Stars A and B, 
suggested as candidates for the progenitor by \citet{vandykprog}, are not 
coincident with the SN recovered in the ACS imaging.  (b) Post-explosion 
ACS/WFC F555W showing SN 1999ev, $m_{V}=23.91\pm0.07$, and associated 
light echo feature not observed in the pre-explosion imaging.  (c) 
Post-explosion ACS/WFC F814W imaging, the SN has magnitude 
$m_{I}=23.35\pm0.08$.{\bf -see jpeg}} 
\label{99evstamp}
\end{figure*}
At the SN location on the pre-explosion frame WF2 chip, with an uncertainty of $\pm0.2$ WF pixels, is a star with 
$m_{F555W}=24.64\pm0.17$, detected at a $4.8\sigma$ level.  This is compared with the new object in the post-explosion ACS imaging with $m_{V}=23.92\pm0.07$. 
Four stars fall within the absolute astrometric error circle on the pre-explosion imaging.  The post-explosion imaging clearly identifies one of these stars, with sub-pixel accuracy, as the progenitor.  \citet{vandykprog} identified two stars, A and B, on the pre-explosion F555W frame and suggested these as candidates for the progenitors.   In addition a separate star C, following the nomenclature of \citet{vandykprog}, is detected $\approx0.2\arcsec$ from the progenitor in the pre-explosion image by both {\sc DAOphot} (at a $3.3\sigma$ level) and HSTphot \citep{dolphhstphot} photometry  with magnitude $m_{F555W}=25.03\pm0.24$.  The post-explosion imaging allows for the confident rejection of stars A, B and C as the progenitor.  A circular feature in the post-explosion imaging, unobserved in the pre-explosion images, is interpreted as a light echo from the SN \citep{liu93jecho}.  This is shown in figure \ref{99evstamp}b.
The brightness of the SN in the B and I bands was: $m_{B}=24.61\pm0.09$ and $m_{I}=23.35\pm0.08$. The NED value for the foreground reddening 
towards NGC 4724 is $E(B-V)=0.02$.  The inclination of the galaxy of $71.8\degr$, the paucity of observed stars in the vicinity of the SN and 
obvious dust lanes imaged in the post-explosion frames suggests a higher 
reddening is appropriate.  Five stars, within $6\arcsec$ of the SN were 
utilised to determine a line of sight reddening of $E(B-V)=0.15\pm0.05$.  This implied an absolute magnitude for the progenitor of $M_{F555W}=-6.5\pm0.3$.
%
\subsection{SN 2000ds}
\label{00dsobssec}
SN 2000ds was discovered by \citet{00dsiauc1} on 2000 October 10.4
in the galaxy of NGC 2768.   The SN was classified as a type-Ib 
several months past maximum.  \citet{vandykprog} give the position of SN 2000ds as $\mathrm{\alpha=9^{h}11^{m}36^{s}.28 ,  \delta=+60\degr01\arcmin 43\arcsec.3}$.
\begin{figure*}
\caption{Pre- and post-explosion of the site of SN 2000ds in NGC 2768. 
(a) Pre-explosion F814W WFPC2 image of the site of SN 2000ds from 1995.  Note the bright peak in the bottom right quadrant is a hot pixel.  
(b) Pre-explosion F555W WFPC2 images of the site of SN 2000ds from 1998. 
(c) Pre-explosion F814W WFPC2 image of the site of SN 2000ds from 1998.  Note the bright peak in the bottom right quadrant is a hot pixel.
(d) Post-explosion F435W ACS/WFC image of the site of SN 2000ds.  Within the positional error circle nothing is recovered - suggesting both the progenitor and SN have not been observed in the pre- and post-explosion imaging. 
(e) Post-explosion F555W ACS/WFC image of the site of SN 2000ds.
(f) Post-explosion F814W ACS/WFC image of the site of SN 2000ds. {\bf - 
see jpeg}}
\label{00dsstamp}
\end{figure*}
Two epochs of pre-explosion observations were available in the HST archive.  
A single WFPC2 F555W observation was made on 1995 December 4, as part of 
program GO-5999 (PI: A. Phillips).  This was an individual frame, without a 
companion frame for the removal of cosmic rays (CRs).  The corresponding F814W frame, composed of two CR-split observations, was only 320s.  The longer F555W and F814W observations from 1998 (program GO-6587, PI: D. Richstone) were utilised to attempt to detect a progenitor.  The post-explosion ACS/WFC images 
were astrometrically recalibrated, with 8 APM 
stars, with a precision of $\pm0.23\arcsec$.  The pre-explosion images from 1998 were aligned with the ACS images with a precision of 
$\pm0.02\arcsec$.  There is no discernible change between the pre-explosion and  post-explosion images.  This indicates that 
the SN has not been recovered in the late time images down to $3\sigma$ 
limiting magnitudes: $m_{F435W}=27.55\pm0.32$, $m_{F555W}=27.10\pm0.32$ 
and 
$m_{F814W}=26.42\pm0.31$.  The progenitor has 
not been detected in the pre-explosion images to limiting magnitudes of 
$m_{F555W}=25.96$ and $m_{F814W}=25.40$.  The foreground 
reddening, quoted by NED towards NGC 2768, is $E(B-V)=0.044$.  The limiting absolute magnitudes of the progenitor were $M_{F555W} \ge -6.02$ and $M_{F814W} \ge -6.52$.
%
\subsection{SN 2000ew}
\label{00ewobssec}
SN 2000ew was discovered by \citet{00ewiauc1} on 2000 November 28.48
in the galaxy of NGC 3810, an Sc galaxy with an inclination
of $47.8\degr$.  \citet{00ewiauc2} classified the SN as a type Ia, although
\citet{00ewiauc3} later reclassified 2000ew as a type Ic SN.  The latest
estimate of the position of SN2000ew is $\mathrm{\alpha=11^{h}40^{m}58^{s}.60 ,
\delta=+11\degr27\arcmin 55\arcsec.8}$ \citep{vandykprog}.  The SN location
had been imaged with WFPC2 with the F606W filter, with an exposure time of
160s, on 1994 November 04 (program GO-5446, PI: G. Illingworth).  This was 6.07 years prior to the discovery
of SN2000ew.  Two sets of post-explosion imaging were acquired with the
WFPC2 and ACS/WFC instruments.  
\begin{figure*}
\caption{Pre- and post-explosion imaging of the site SN 2000ew in the 
galaxy NGC 3810. (a) Pre-explosion WFPC2 F606W image, the SN fell on WF4 
chip.  At the SN location no object is seen.  The WFPC2 images have been 
scaled, rotated and aligned with post-explosion ACS/WFC observations.  (b) 
Post-explosion WFPC2 F814W image, with the SN on the WF3 chip, acquired as 
part of program HST GO 9042.  (c) Post-explosion ACS/WFC F555W imaging 
with the fading SN.{\bf -see jpeg}} 
\label{00ewstamp}
\end{figure*}
WFPC2 images of the SN, 0.94 years after
discovery, were acquired as part of program GO 9042 (PI: S. Smartt).  Imaging of the SN,
and host galaxy, was conducted in the F450W and F814W for 460s each.
Observations of the SN, 1.58 years after discovery, were made with the
ACS/WFC in F435W, F555W and F814W bands.  The ACS/WFC images were
astrometrically recalibrated with four stars from the APM catalogue.
The SN position was located on the ACS images with a precision of
$0.5\arcsec$.  The pre- and post-explosion WFPC2 images were aligned with
the ACS images, with a precision of $0.025\arcsec$ and $0.03\arcsec$
respectively.  At the SN location the pre-explosion F606W frame and the
post-explosion F555W ACS frame appear similar.  In the post-explosion
WFPC2 frames, however, a new object is seen and this is identified as the
SN.  We measure the brightness of this object in the F450W and F814W
filters as $m_{F450W}=22.19\pm0.08$ and $m_{F814W}=20.94\pm0.03$.  Nothing is detected at the SN location on the pre-explosion WFPC2 F606W frame, with a $3\sigma$ detection limit of $m_{F606W}=24.6$.  The reddening determined from the neighbouring stars is
$E(B-V)=0.01\pm0.03$.  We adopt, therefore, the NED value for the foreground
reddening of $E(B-V)=0.044$ towards to SN 2000ew.  This implied a limiting absolute magnitude for the progenitor of $M_{F606W} \ge -6.42$.
%
\subsection{SN 2001B}
\label{01bobssec}
SN 2001B was discovered by \citet{01biauc1} in the Sc galaxy IC 391. While 
\citet{01biauc2} classified SN 2001B as a type-Ia, \citet{01biauc3} later 
reclassified the SN as type Ib $\sim7$ days post-maximum.  
\citet{01biauc1} reported a position for the SN of 
$\mathrm{\alpha=4^{h}57^{m}19^{s}.24 , \delta=+78\degr11\arcmin 16\arcsec.5}$, 
whereas observations by \citet{vandykprog} gave a position 
$\mathrm{\alpha=4^{h}57^{m}19^{s}.24 , \delta=+78\degr11\arcmin 16\arcsec.6}$ with an 
uncertainty of $\pm0.2\arcsec$.  An individual 70s WFPC2 F555W image was 
available in the HST archive, acquired as part of program GO-5104 (PI: 
J. Westphal).  This frame did not have a companion observation for the 
rejection of cosmic rays and, hence, in this case 
photometry was conducted with care to avoid inclusion of cosmic rays.  
This image was acquired prior to the "cool down" of the WFPC2 instrument, and 
the CTE appropriate for the epoch of acquisition of this frame has not 
been formalised \citep{dolp00cte}.  The 
post-explosion ACS/WFC F555W was astrometrically recalibrated using 8 APM 
stars with a precision of $\pm0.34\arcsec$.  
The post-explosion images were aligned with the single pre-explosion frame 
to within $\pm0.04\arcsec$.  The SN location was determined to be on the 
WF3 chip on the pre-explosion frame.  The pre- and post-explosion images 
are shown as figure \ref{01bstamp}.  
\begin{figure*}
\caption{Pre- and post-explosion imaging of the site of SN 2001B in IC 391.  
(a) Pre-explosion WFPC2 F555W imaging.  The SN location was determined to be 
on the WF3 chip.  Star A is observed in both the pre- and post-explosion 
images, although \citet{vandykprog} identified it as the likely 
progenitor.  
(b) Post-explosion F555W imaging of SN 2001B.  The SN was 
recovered in this late imaging.  At the same location on the pre-explosion 
image no object is detected.  (c) Post-explosion ACS/WFC F814W imaging of 
SN 2001B. {\bf - see jpeg}}
\label{01bstamp}
\end{figure*}
\citet{vandykprog} identified star A, of figure \ref{01bstamp}, as the 
likely progenitor star.  Inspection of the pre- and post-explosion frames 
shows, however, a new object on the post-explosion image within the 
astrometric error circle for the SN.  No object is detected at this 
position on the pre-explosion image to a 
limit of $m_{F555W}=24.301\pm0.145$.  We identify this new 
object 
on the post-explosion frame as SN 2001B. 
The SN 
brightness in the ACS/WFC imaging was determined to be: 
$m_{B}=23.52\pm0.02$, $m_{V}=22.92\pm0.02$ and $m_{I}=22.46\pm0.02$.  The 
reddening towards SN 2001B was determined, using 68 stars within $6\arcsec$ 
of the SN, to be $E(B-V)=0.102\pm0.030$.  This is in agreement with NED value 
for the fore-ground reddening of $E(B-V)=0.127$.  The absolute magnitude for the progenitor on the F555W pre-explosion imaging was $M_{F555W} \ge -8.01$.
\section{Discussion}
\label{disc}
The detection limits of the progenitors, presented in \S\ref{obssec}, were interpreted as mass limits by placing the detection thresholds on the Hertzprung-Russell (HR) diagram.  Mass limits were determined by comparing regions excluded by the detections limits, in which a progenitor would have been observed, with the progenitors predicted by stellar evolution models.  Here we adopt the Geneva group non-rotating stellar evolution models of \citet{lej01}\footnote{http://webast.ast.obs-mip.fr/stellar}.  \citet{drill00} provide the standard colours of supergiants.  These colours were transformed using the relations of \citet{dolp00cte} to calculate the colour correction, as a function of supergiant spectral type, between the magnitudes in the WFPC2 photometric system and the corresponding Johnson-Cousins magnitude.  The bolometric corrections, quoted by \citet{drill00}, were then utilised to calculate a luminosity for each spectral type.  The colour and bolometric corrections and temperatures are given in table \ref{magtab}.
\begin{table*}
\caption{A table showing the colours and bolometric corrections of supergiants, as given by \citet{drill00}.  The corrections between magnitudes in the WFPC2 photometric system and the corresponding standard Johnson-Cousins magnitude were calculated using the updated transformations of \citet{dolp00cte}.}
\label{magtab}
\begin{tabular}{lrrrrrrrrrrrrr}
\hline\hline
Sp &  M(V) &  B-V  &  U-B  &  V-R  &  R-I  &  $\mathrm{T_{eff}}$ &  BC    &  V-I  & B-F439W & B-F450W & V-F555W & V-F606W &  I-F814W \\
\hline
O9 &  -6.5 &  -0.27&  -1.13&  -0.15&  -0.32&  32000&  -3.18 &  -0.47&  -0.007 &  -0.060 &  0.030  &  -0.108 &  0.035   \\
B2 &  -6.4 &  -0.17&  -0.93&  -0.05&  -0.15&  17600&  -1.58 &  -0.2 &  -0.003 &  -0.037 &  0.011  &  -0.041 &  0.013  \\
B5 &  -6.2 &  -0.1 &  -0.72&  0.02 &  -0.07&  13600&  -0.95 &  -0.05&  -0.001 &  -0.021 &  0.003  &  -0.004 &  0.003  \\
B8 &  -6.2 &  -0.03&  -0.55&  0.02 &  0    &  11100&  -0.66 &  0.02 &   0.000 &  -0.005 &  -0.001 &   0.014 &  -0.001  \\
A0 &  -6.3 &  -0.01&  -0.38&  0.03 &  0.05 &  9980 &  -0.41 &  0.08 &  0.000  &  0.000  &  -0.004 &  0.029  &  -0.005  \\
A2 &  -6.5 &  0.03 &  -0.25&  0.07 &  0.07 &  9380 &  -0.28 &  0.14 &  0.000  &  0.009  &  -0.007 &  0.045  &  -0.008 \\
A5 &  -6.6 &  0.09 &  -0.08&  0.12 &  0.13 &  8610 &  -0.13 &  0.25 &  0.000  &  0.023  &  -0.011 &  0.073  &  -0.014  \\
F0 &  -6.6 &  0.17 &  0.15 &  0.21 &  0.2  &  7460 &  -0.01 &  0.41 &  -0.002 &  0.041  &  -0.017 &  0.115  &  -0.021  \\
F2 &  -6.6 &  0.23 &  0.18 &  0.26 &  0.21 &  7030 &  0     &  0.47 &  -0.004 &  0.055  &  -0.018 &  0.131  &  -0.024  \\
F5 &  -6.6 &  0.32 &  0.27 &  0.35 &  0.23 &  6370 &  -0.03 &  0.58 &  -0.008 &  0.075  &  -0.021 &  0.160  &  -0.028  \\
F8 &  -6.5 &  0.56 &  0.41 &  0.45 &  0.27 &  5750 &  -0.09 &  0.72 &  -0.026 &  0.130  &  -0.023 &  0.198  &  -0.032  \\
G0 &  -6.4 &  0.76 &  0.52 &  0.51 &  0.33 &  5370 &  -0.15 &  0.84 &  -0.049 &  0.175  &  -0.025 &  0.231  &  -0.034  \\
G2 &  -6.3 &  0.87 &  0.63 &  0.58 &  0.4  &  5190 &  -0.21 &  0.98 &  -0.064 &  0.200  &  -0.025 &  0.269  &  -0.037   \\
G5 &  -6.2 &  1.02 &  0.83 &  0.67 &  0.44 &  4930 &  -0.33 &  1.11 &  -0.088 &  0.233  &  -0.024 &  0.306  &  -0.038   \\
G8 &  -6.1 &  1.14 &  1.07 &  0.69 &  0.46 &  4700 &  -0.42 &  1.15 &  -0.111 &  0.260  &  -0.024 &  0.317  &  -0.038  \\
K0 &  -6   &  1.25 &  1.17 &  0.76 &  0.48 &  4550 &  -0.5  &  1.24 &  -0.134 &  0.285  &  -0.023 &  0.342  &  -0.038  \\
K2 &  -5.9 &  1.36 &  1.32 &  0.85 &  0.55 &  4310 &  -0.61 &  1.4  &  -0.159 &  0.309  &  -0.020 &  0.388  &  -0.038 \\
K5 &  -5.8 &  1.6  &  1.8  &  1.2  &  0.9  &  3990 &  -1.01 &  2.1  &  -0.220 &  0.362  &  0.010  &  0.567  &  -0.020 \\
M0 &  -5.6 &  1.67 &  1.9  &  1.23 &  0.94 &  3620 &  -1.29 &  2.17 &  -0.240 &  0.378  &  0.014  &  0.569  &  -0.017  \\
M2 &  -5.6 &  1.71 &  1.95 &  1.34 &  1.1  &  3370 &  -1.62 &  2.44 &  -0.252 &  0.387  &  0.034  &  0.583  &  -0.002  \\
M5 &  -5.6 &  1.8  &  1.6  &  2.18 &  1.96 &  2880 &  -3.47 &  4.14 &  -0.280 &  0.406  &  0.247  &  0.890  &  0.172   \\
\hline\hline
\end{tabular}
\end{table*}
In addition the Potsdam Wolf-Rayet (WR) Star Models synthetic spectra \citep{wolfspec}\footnote{http://www.astro.physik.uni-potsdam.de/ \\ $\sim$wrh/PoWR/powrgrid1.html} were utilised to explore the luminosity behaviour of the detection threshold for WR stars.  The synthetic spectra used are for an approximately solar abundance.  These models are dependent, however, on two parameters: the effective temperature and the Transformed Radius, $R_{t}$.  Bolometric corrections and colour corrections (between the Johnson V band and the WFPC2 F555W and F606W bands) were calculated from the synthetic spectra using the STSDAS {\it Synphot} package.  Models were selected from the grid of synthetic spectra to form three groups with the maximum, median and minimum $R_{t}$ for a given effective temperature.  This takes into the account the spread in luminosities of WR stars, as parameterised by $R_{t}$.  The results from these models are presented in table \ref{wolftab}.
\begin{table}
\caption{The colour corrections, between the WFPC2 and Johnson-Cousins photometric systems, and bolometric corrections for a range of WR star parameters calculated from the synthetic spectra of \citet{wolfspec}.   The table is separated by the maximum, median and minimum Transformed Radii $R_{t}$ at each effective temperature.}
\label{wolftab}
Minimum\\
\begin{tabular}{rrrrrr}
\hline\hline
Temp&Model&$\log R_t$&V-F555W&V-F606W&$B.C.$ \\
(K)& No. & & & & \\
\hline
70800&10-21& 0 &0.053&0.103&-2.737\\
63100&09-20&0.1&0.049&0.106&-2.700\\
56200&08-19&0.2&0.044&0.089&-2.667\\
50100&07-18&0.3&0.040&0.079&-2.632\\
44700&06-17&0.4&0.033&0.070&-2.585\\
39800&05-16&0.5&0.028&0.057&-2.527\\
35500&04-15&0.8&0.024&0.052&-2.441\\
31600&03-10&1.1&0.024&0.008&-2.751\\
\hline\hline
\end{tabular}
Median\\
\begin{tabular}{rrrrrr}
\hline\hline
Temp&Model&$\log R_t$&V-F555W&V-F606W&$B.C.$ \\
(K)& No. & & & & \\
\hline
177800&18-20&0.1&0.165&0.068&-5.974\\
158500&17-19&0.2&0.191&0.075&-5.850\\
141300&16-18&0.3&0.195&0.070&-5.739\\
125900&15-17&0.4&0.195&0.061&-5.642\\
100000&13-16&0.5&0.171&0.055&-5.192\\
89100&12-15&0.6&0.166&0.056&-5.133\\
70800&10-13&0.8&0.121&0.033&-4.796\\
63100&09-12&0.9&0.104&0.020&-4.622\\
56200&08-11&1.0 &0.102&0.020&-4.350\\
50100&07-10&1.1&0.099&0.088&-4.083\\
44700&06-09&1.2 &0.096&0.154&-3.788\\ 
39800&05-08&1.3 &0.051&0.103&-3.509\\
35500&04-05&1.6&0.028&0.018&-3.202\\
\hline\hline
\end{tabular}
Maximum\\
\begin{tabular}{rrrrrr}
\hline\hline
Temp&Model&$\log R_t$&V-F555W&V-F606W&$B.C.$ \\
(K)& No. & & & & \\
\hline
199500&19-16&0.5&0.064&0.012&-7.818\\
177800&18-13&0.8&0.046&-0.025&-7.898\\
158500&17-12&0.9&0.043&-0.033&-7.561\\
141300&16-11&1.0&0.041&-0.038&-7.227\\
125900&15-10&1.1&0.041&-0.046&-7.016\\
112200&14-09&1.2&0.039&-0.041&-6.633\\
100000&13-08&1.3&0.044&-0.042&-6.406\\
79400&11-06&1.5&0.040&-0.049&-5.804\\
70800&10-05&1.6&0.034&-0.052&-5.431\\
56200&08-04&1.7&0.032&-0.045&-4.707\\
\hline\hline
\end{tabular}
\end{table}
The detection thresholds were converted to absolute magnitudes, using the 
extinctions determined here and the distance moduli quoted by LEDA (given in table \ref{Snhosts}).  
\begin{equation}
m_{X;WFPC2}-M_{X;WFPC2}=5logd-5+A_{X;WFPC2}
\end{equation}
for the general filter $X$ in the WFPC2 photometric system.\\
The bolometric magnitudes, as a function of the effective temperatures for each spectral type,  were then calculated from the absolute magnitudes using the formula:
\begin{eqnarray}
M_{bol}(T_{eff}) & = & M_{X;WFPCC2}+(M_{X;J-C}-M_{X;WFPC2})\nonumber \\
                 &   & +(M_{V}-M_{X;J-C})+B.C. 
\end{eqnarray}
Where $M_{X;J-C}-M_{X;WFPC2}$ is the colour correction for the general passband $X$ in the WFPC2 photometric system and the corresponding passband in the Johnson-Cousins photometric system.  The luminosity was calculated using the standard formula: 
\begin{equation}
log(L(T_{eff})/L_{\odot})=\frac{M_{bol}-4.74}{-2.5}
\end{equation}
In this way the luminosity limit was calculated for a range of temperatures over all supergiant spectral types.  The limiting luminosity describes an ``arc'' across the HR diagram depending on the colour and temperature (and hence the bolometric correction).  A progenitor star with a luminosity greater than the determined limit would have been observed in the pre-explosion frames.  In addition the stellar populations surrounding the progenitors were examined.  The approximate ages of these stars were determined by examining the locus of these stars on colour-magnitude diagrams.  Isochrones, produced by the Geneva stellar evolution group \citep{lej01}, were overlaid on these diagrams to determine the ages.  Metallicities were estimated, in the broad groups of SMC, LMC, Solar and twice Solar, by considering the relationship of \citet{metapil} and the absolute magnitudes of the host galaxies quoted in LEDA.  In addition the metallicity gradient of NGC 2403 \citep{meta2403} was adopted to estimate the metallicity at the radii of the SNe.
\subsection{SN 1999an}
\label{99ansec}
\begin{figure}
\rotatebox{-90}{\includegraphics[width=6cm]{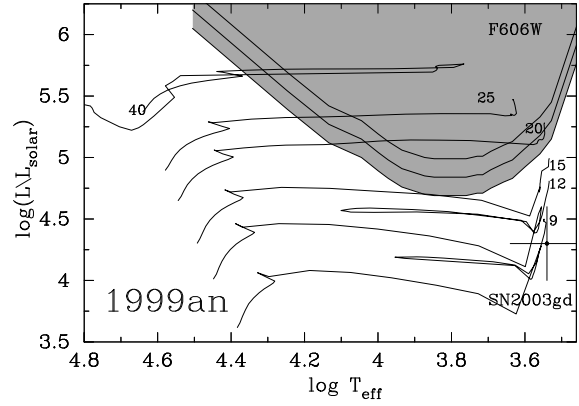}}
\caption{An HR diagram showing the luminosity and temperature region (in 
grey) where stars would have been detected in pre-explosion imaging.  The 
progenitor of SN 1999an is not detected to an absolute magnitude of $M_{F606W} \ge -7.5\pm0.3$ (or $\Delta L = \pm 0.12$).  The detection region shows that the progenitor of SN 1999an had a mass $M_{ZAMS} < 20M_{\odot} $ for it not to have been detected in the HST WFPC2 pre-explosion F606W imaging.}
\label{99anhr}
\end{figure}

\begin{figure}
\rotatebox{-90}{\includegraphics[width=6cm]{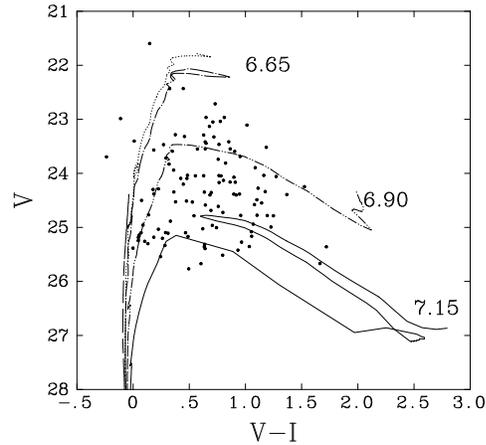}}
\caption{Colour-magnitude diagram showing the locus of stars in the vicinity of the progenitor of SN 1999an.  Overlaid are solar metallicity isochrones shifted to the distance, extinction and reddening determined for the progenitor in IC 755.  The age of the stellar population is $\mathrm{log(Age/years)=7.15\pm0.25}$.  This age is consistent with the expected life time for stars with $M_{ZAMS}=15-20M_{\odot}$.  The large spread in apparent ages is due to the varying depth, reddening and ages being probed in nearby lines of sight because of the inclination of IC 755.}
\label{99aniso}
\end{figure}

The detection threshold $M_{F606W} \ge -7.5\pm0.3$ for the progenitor of SN 1999an was plotted on an HR diagram.   The metallicity was assumed to be solar.  The determination of the appropriate metallicity for SN 1999an and its progenitor was hampered by the high inclination of the host galaxy and the line of sight proximity of the SN to the centre of the galaxy.  An HR diagram, showing the region excluded by the pre-explosion imaging and the non-detection of the progenitor, is shown as figure \ref{99anhr}.  A red supergiant progenitor, with an initial mass $>20M_{\odot}$, would have been detected in the pre-explosion F606W image. 
The age of the surrounding stellar population, see figure \ref{99aniso}, is $\approx$14Myr.  This is consistent the approximate life time of stars with $M_{ZAMS}=15-20M_{\odot}$.  The inclination of IC 755 complicates the age determination procedure by probing various radii of the host galaxy, as well as reddening and ages in nearby lines of sight.  This is responsible for the apparent spread in the observed colour-magnitude diagram for stars in IC 755.  This mass limit is consistent with the mass limits determined for type II-P SNe \citep{smartt01du}.
\subsection{SN 1999br}
\label{99brsec}
\begin{figure}
 \rotatebox{-90}{\includegraphics[width=6cm]{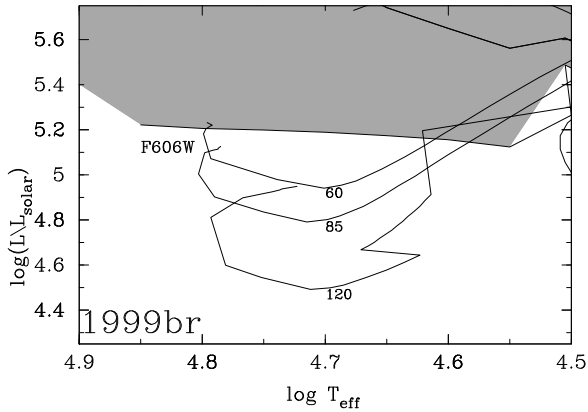}}
\caption{A figure showing the detection limit of the F606W pre-explosion observation, at the location of SN 1999br, on the blue side of the HR diagram.  Overlaid are solar metallicity stellar evolution tracks for $M_{ZAMS}=60$, $85$ and $120M_{\odot}$.}
\label{99brhrblue}
\end{figure}

\begin{figure}
\rotatebox{-90}{\includegraphics[width=6cm]{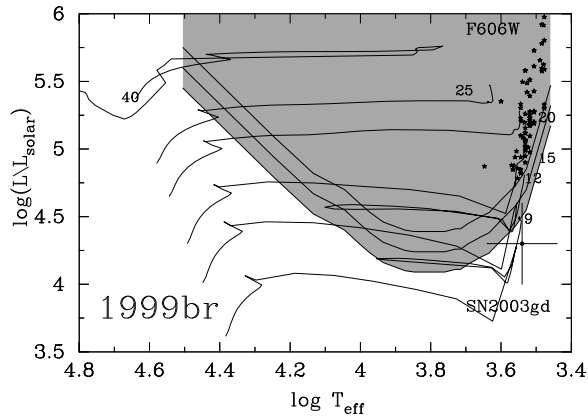}}
\caption{A figure showing the detection limit of the F606W pre-explosion 
observation, of the location of SN 1999br, on the red side of the HR 
diagram.  Stars with $12M_{\odot}<M_{ZAMS}<25M_{\odot}$ are excluded as 
progenitors in a single star scenario. The locus of observed red supergiants, in the LMC \citep{masslmc03}, is shown by the starred points.  The brightest of these red supergiants would have been detected in the pre-explosion imaging.}
\label{99brhrred}
\end{figure}

The metallicity at the position of SN1999br, in NGC 4900, was estimated as 
approximately solar.   Exclusion zones were plotted on the HR diagram for the limiting detection magnitude of the pre-explosion F606W image.  These are shown as figures \ref{99brhrblue} and \ref{99brhrred}.  A useful upper mass limit cannot be placed on the progenitor with the F606W observation and considering a WR star progenitor.  \citet{past99br} find, from observations of the light curves, a long plateau and a low tail luminosity indicates a massive H envelope and low Ni production.   The long light curve plateau is consistent with a red supergiant progenitor, which the F606W observation limits to $M_{ZAMS}\le 12M_{\odot}$.  \citet{tur97D} estimate a progenitor mass of $26M_{\odot}$ for the similarly faint SN 1997D.  In this scenario the low luminosity of the SN arises from the fall back of Ni onto a newly formed black hole.  Models of \citet{heg03} suggests this happens for stars with $30M_{\odot} \le M_{ZAMS} \le 40M_{\odot}$.
 Alternatively \citet{chug00} suggest a low mass progenitor, $8-12M_{\odot}$, for SN 1997D.  The latter estimate is a similar mass to the observed progenitor of the normal type II-P SNe 1999em, 1999gi, 2001du and 2003gd \citep{smartt03gd}.  
\citet{zamp99br} and \citet{past99br} estimate, from models of photometric and spectroscopic observations of SN 1999br, an ejected envelope mass of $14-19M_{\odot}$ from an intermediate mass progenitor of mass $\ge16M_{\odot}$.\\
The pre-explosion observation, however, rules out a bright red supergiant progenitor for SN 1999br, as has been proposed for SN 1997D.  The locus of observed red supergiants, in the LMC, on the HR diagram is shown on figure \ref{99brhrred}.  A very massive red supergiant progenitor would have been significantly detected on the pre-explosion F606W image.
\begin{figure}
\rotatebox{-90}{\includegraphics[width=6cm]{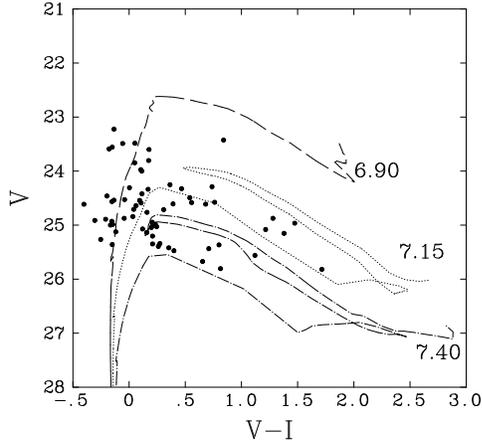}}
\caption{A V, V-I colour magnitude diagram for stars within $6\arcsec$ of the site of SN 1999br.  The surrounding population defines a tight locus on the diagram.  Overlaid are solar metallicity isochrones, corrected for the distance of and reddening towards NGC 4900.  The surrounding stellar population has an age of $\mathrm{log(Age/years)=7.15\pm0.25}$.  This age is consistent with the lifetime of a star with $M_{ZAMS}\sim9-20M_{\odot}$}.
\label{99briso}
\end{figure}
The pre-explosion observation supports the hypothesis of \citet{chug00}
that such faint SNe arise from low mass progenitors.  The simultaneous
production of both faint and normal type II-P SNe from the same range
of initial masses  $M_{ZAMS}=8-12M_{\odot}$ is not unexpected.
The evolution of stars in this range is particularly sensitive
to initial mass as it
includes the transition between electron-degenerate and non-degenerate
stellar cores and slight differences in initial mass will lead
to different SN outcomes \citep{nom84,eld04}.  SNe originating from the
core-collapse of degenerate cores are predicted to produce little Ni due
to the presence of steep density gradients at the edges of the cores.
This leads to only small amounts of material being sufficiently heated to
produce Ni \citep{heg03}.
The age of the progenitor was estimated from the approximate ages of nearby stars (within $6\arcsec$ of the SN location) in NGC 4900.  The locus of these stars on the V, V-I colour magnitude diagram is shown as figure \ref{99briso}.  The age was determined as  $\mathrm{log(Age/years)=7.15\pm0.25}$, which is consistent with a progenitor of initial mass $\sim9-20M_{\odot}$.
\subsection{SN 1999ev}
\label{99evsec}
\begin{figure}
\rotatebox{-90}{\includegraphics[width=6cm]{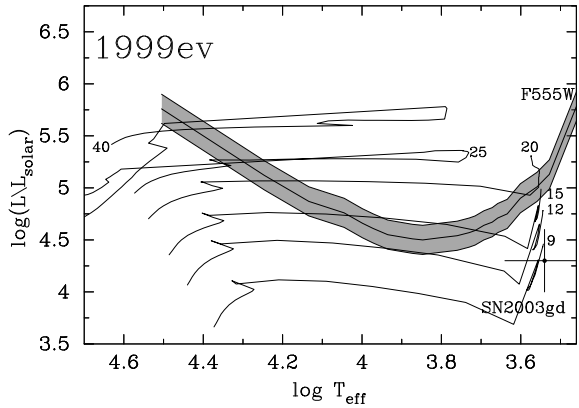}}
\caption{An HR diagram showing the locus of allowed luminosities for the 
progenitor of SN 1999ev, given an absolute magnitude of $M_{F555W}=-6.5$.  
The thick line shows the median luminosity, at a given temperature, for 
that absolute magnitude.  The two thin lines show the upper and lower 
boundaries of the luminosity given the uncertainty in the absolute magnitude, the distance to the host galaxy and the reddening along the line of sight ($\Delta L=\pm0.11$).  The observed progenitor, assuming a red supergiant, had mass $M_{ZAMS}=15-18_{\odot}$.}
\label{99evhr}
\end{figure}

\begin{figure}
\rotatebox{-90}{\includegraphics[width=6cm]{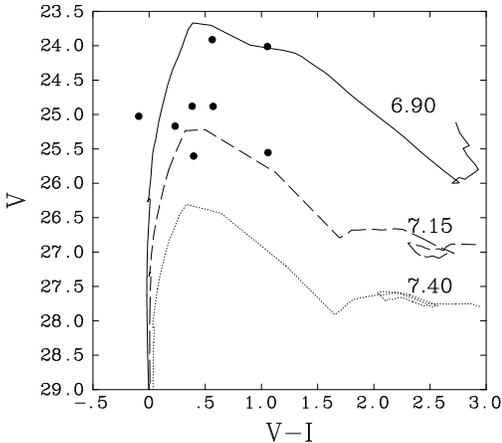}}
\caption{Colour-magnitude diagram showing the locus of stars in the vicinity of the progenitor of SN 1999ev.  Overlaid are twice-solar metallicity isochrones shifted to the distance, extinction and reddening determined for the progenitor.  The stellar population is clustered around the isochrones for $\mathrm{log(Age/years)=7.15\pm0.25}$.  This age is consistent with the expected life time for stars with $M_{ZAMS}=15-20M_{\odot}$.}
\label{99eviso}
\end{figure}

The brightness of the identified progenitor star for SN1999ev was plotted on an HR diagram, shown as figure \ref{99evhr}.   The proximity of the SN to the centre of NGC 4724 suggests that the appropriate metallicity for the progenitor of SN 1999ev is closer to twice solar.  A study of the metallicity of NGC 4274 and its gradient has not been conducted.  Because the progenitor star was only observed in a single filter the absence of any colour information gives a degeneracy in temperature. 
The allowed region on the HR diagram for the progenitor of SN encompasses the red supergiant end points for stars with $15<M_{ZAMS}<18M_{\odot}$.
The photometry of the stars surrounding the progenitor is shown of the colour-magnitude diagram figure \ref{99eviso}.  The age of these stars, $\mathrm{log(Age/years)=7.15\pm0.25}$, is consistent with the theoretically predicted life times for stars with $M_{ZAMS}=15-20M_{\odot}$ (6.95 and 7.1 respectively).
This mass is higher than both the determined progenitor mass for SN 2003gd and the mass limits for a number of normal type II-P SNe by \citet{smartt01du} ($M_{ZAMS}<15M_{\odot}$).  
\subsection{SN 2000ds}
\label{00dssec}
\begin{figure}
\rotatebox{-90}{\includegraphics[width=6cm]{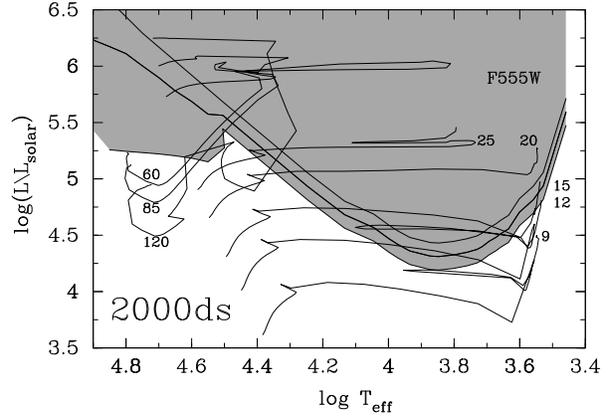}}
\caption{A figure showing the detection threshold of pre-explosion F555W 
WFPC2 imaging at the site of SN 2000ds.   Over plotted are solar 
metallicity stellar evolution tracks.   The end of the $40M_{\odot}$ track 
lies at the same position as the end of the $85M_{\odot}$ track.  The F555W 
detection limit excludes stars with $12M_{\odot}<M_{ZAMS}<25M_{\odot}$.}
\label{00dshrf555w}
\end{figure}

\begin{figure}
\rotatebox{-90}{\includegraphics[width=6cm]{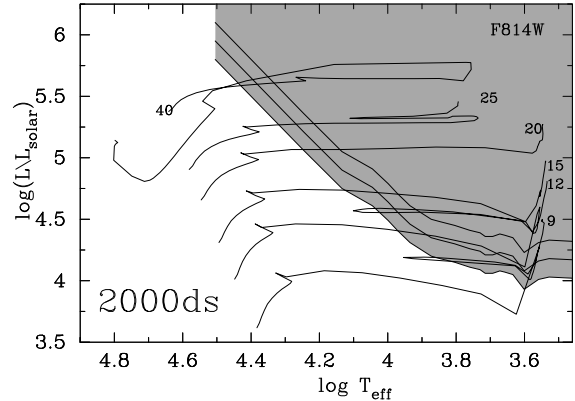}}
\caption{A figure showing the detection threshold of pre-explosion F814W 
WFPC2 imaging of the site of SN 2000ds.   Over plotted are solar 
metallicity stellar evolution tracks.  The F814W detection limit excludes 
a red supergiant progenitor with $M_{ZAMS}>7M_{\odot}$.}
\label{00dshrf814w}
\end{figure}

\begin{figure}
\rotatebox{-90}{\includegraphics[width=6cm]{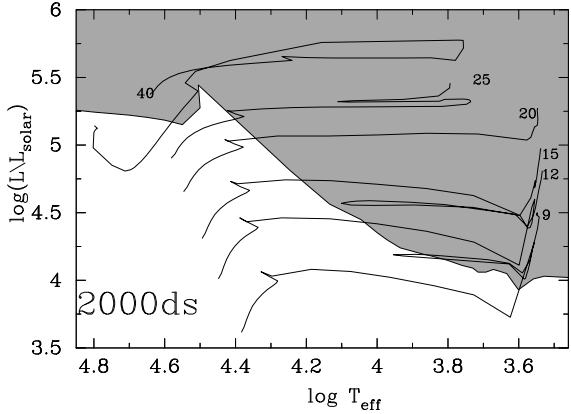}}
\caption{An HR diagram simultaneously showing the detection thresholds from all pre-explosion imaging of SN 2000ds.  A red progenitor is excluded by the deep pre-explosion F814W imaging.}
\label{00dshrtot}
\end{figure}

The two colour detection limits from the pre-explosion F555W and F814W were simultaneously utilised to place a mass limit on the progenitor of SN 2000ds.  The classification of the host galaxy makes the determination of the appropriate metallicity and the interpretation of the progenitor difficult.  \citet{van03}  classify the host galaxy NGC 2768 as E3/Sa and the inclination for the galaxy, quoted by LEDA, is unclear.  In this case, therefore, a solar metallicity has been adopted.  The detection thresholds from the pre-explosion F555W and F814W imaging are shown as figures \ref{00dshrf555w} and \ref{00dshrf814w} respectively.  The combined detection limit of both sets of imaging, shown as figure \ref{00dshrtot} exclude a red supergiant progenitor down to $M_{ZAMS}\approx7M_{\odot}$.  A massive blue progenitor cannot be excluded.  If NGC 2768 is an E3 galaxy a single massive progenitor is unlikely and the core-collapse of a blue low mass progenitor in a binary would be a plausible progenitor scenario.
\subsection{SN 2000ew}
\label{00ewsec}
\begin{figure}
\rotatebox{-90}{\includegraphics[width=6cm]{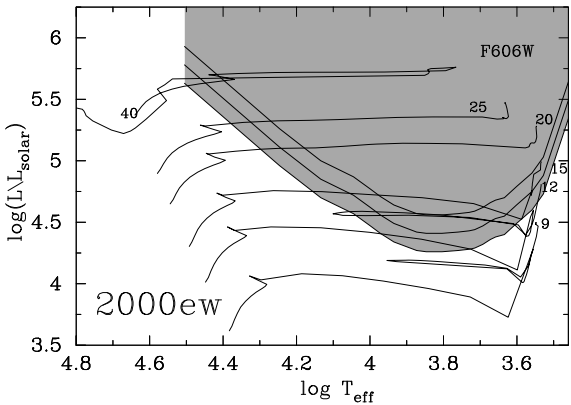}}
\caption{An HR diagram showing the detection threshold for pre-explosion F606W image.  A red supergiant progenitor with an initial mass $>12M_{\odot}$ would have been detected in the pre-explosion imaging.  Overlaid are stellar evolution tracks for $\mathrm{Z=2Z_{\odot}}$.}
\label{00ewhr}
\end{figure}

\begin{figure}
\rotatebox{-90}{\includegraphics[width=6cm]{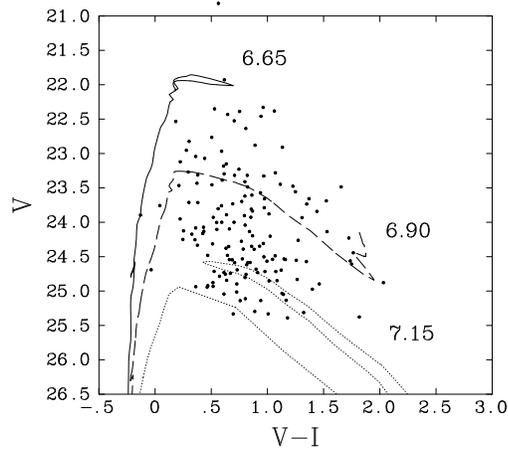}}
\caption{Colour-magnitude diagram showing the locus of stars within $6\arcsec$ of SN 2000ew.  Overlaid are twice solar metallicity isochrones, corrected for the distance and reddening to the host galaxy NGC 3810.  The age of the stellar population is $\mathrm{log(Age/years)=6.9\pm0.25}$, approximately consistent with a progenitor $12M_{\odot}<M_{ZAMS}<40M_{\odot}$.}
\label{00ewiso}
\end{figure}

Given the proximity of the site of SN 2000ew to the centre of the host galaxy NGC 3810 the metallicity has been estimated as twice-solar. 
 The detection threshold for the progenitor of SN 2000ew, from pre-explosion F606W imaging, is shown as figure \ref{00ewhr}.
A massive Wolf-Rayet progenitor ($M_{ZAMS}>25M_{\odot}$) cannot be explicitly excluded from the pre-explosion imaging.  The age of stars in the locality of SN 2000ew, shown as figure \ref{00ewiso}, was determined to be $\mathrm{log(Age)=6.9\pm0.25}$.  This is consistent with the lifetime of stars with $12M_{\odot}<M_{ZAMS}<40M_{\odot}$.  Post-explosion ACS imaging of SN 2000ew was useful to show not only the location of the SN but, by comparison with post-explosion WFPC2 imaging acquired at an early epoch, the SN fading.
\subsection{SN 2001B}
\label{01bsec}
\begin{figure}
\rotatebox{-90}{\includegraphics[width=6cm]{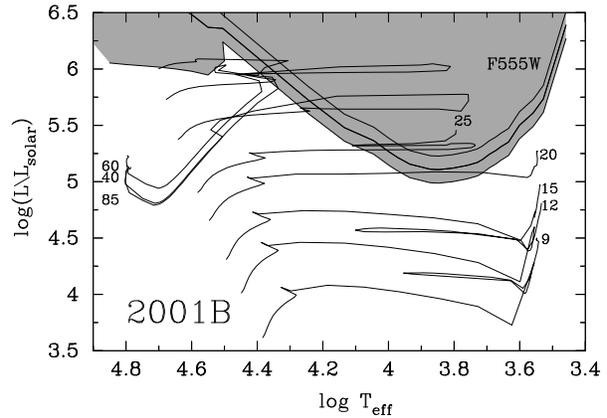}}
\caption{An HR diagram showing the detection threshold for pre-explosion 
F555W imaging of the location of SN 2001B.  The detection threshold allows 
a single red progenitor up to a mass of $M_{ZAMS}=22M_{\odot}$.  A blue 
Wolf-Rayet progenitor is not excluded.}
\label{01Bhr}
\end{figure}

The detection limit for the F555W pre-explosion imaging is shown as figure \ref{01Bhr}.  Solar metallicity stellar evolution tracks were utilised to estimate the mass limits for the progenitor.  The mass limit for a red supergiant progenitor is $M_{ZAMS}<25M_{\odot}$.  Similarly to SN 2000ew a single blue massive progenitor ($>25M_{\odot}$) is also permitted.  A WR progenitor is favoured given the type Ib classification.   
\begin{figure}
\rotatebox{-90}{\includegraphics[width=6cm]{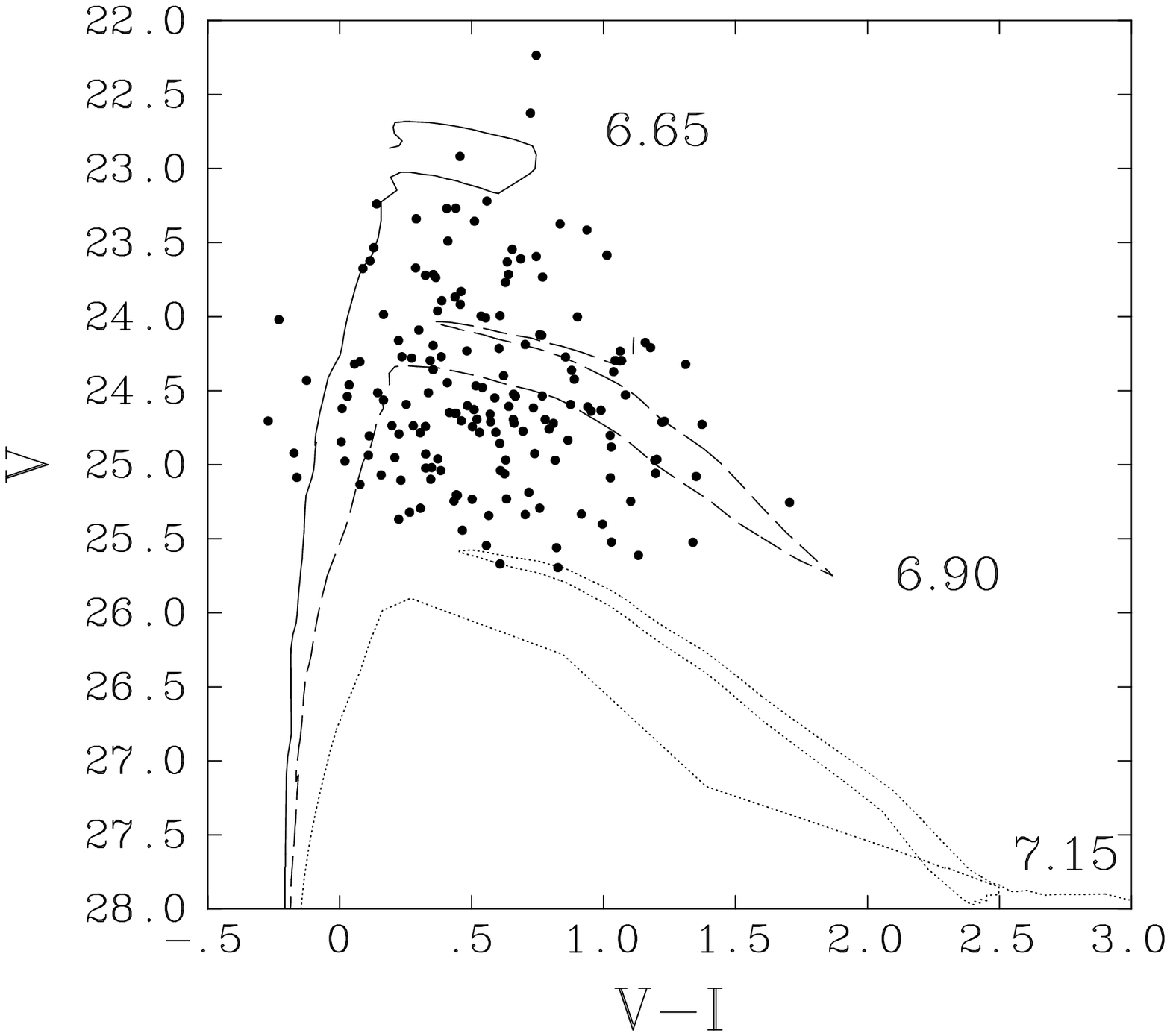}}
\caption{A colour-magnitude diagram showing the locus of stars within $6\arcsec$ of SN 2001B.  Overlaid are solar metallicity isochrones, corrected for the distance and reddening to the host galaxy IC 391.  The age of the stellar population is $\mathrm{log(Age/years)=6.9\pm0.25}$, approximately consistent with a progenitor $12M_{\odot}<M_{ZAMS}<25M_{\odot}$.}
\label{01Biso}
\end{figure}
\subsection{General Discussion}
\begin{table}
\caption{A table summarising the mass limits for the progenitors of SNe 1999an, 1999br, 1999ev, 2000ds, 2000ew and 2001B, as derived from pre-explosion HST WFPC2 imaging.}
\label{snsumm}
\begin{tabular}{lcc}
\hline\hline
SN & Type & Mass Limit \\
\hline
1999an & II(P) &$M_{ZAMS}\le20M_{\odot}$  \\
1999br & IIpec &$M_{ZAMS} \le 12M_{\odot}$   \\
1999ev & II(P) &$15 \le M_{ZAMS} \le 18M_{\odot}$   \\
2000ds & Ib &$M_{ZAMS} > 25M_{\odot}$ or binary?  \\
2000ew & Ic &$M_{ZAMS} > 25M_{\odot}$ (WR?)  \\
2001B  & Ib &$M_{ZAMS} > 25M_{\odot}$ (WR?)    \\
\hline
\end{tabular}
\end{table}
\begin{figure*}
\rotatebox{-90}{\includegraphics[width=10cm]{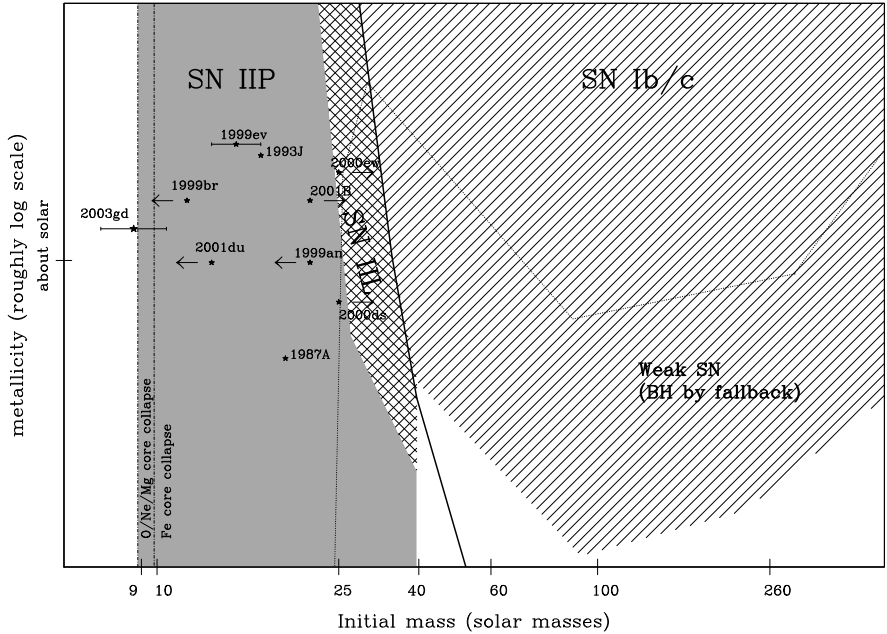}}
\caption{A SN Populations diagram \citep{heg03} showing the locus of the 6 progenitors studied here: 1999an, 1999br, 1999ev, 2000ds, 2000ew and 2001B.  In addition the positions of the three SNe with known progenitors, 1987A, 1993J and 2003gd, are shown.  The upper mass limit for the progenitor of SN 2001du illustrates the expected mass regime for common type II-P SNe.  Each of the mass limits is consistent with the predicted progenitors for each of the SN types.  Approximate metallicities have been adopted to place the progenitors on the diagram.}
\label{snpops}
\end{figure*}
The mass limits for the progenitors of the SNe discussed here are summarised in table \ref{snsumm}.  The mass limits are  presented on a SN populations diagram \citep{heg03}, shown as figure \ref{snpops}.  \citet{smartt01du} present masses or mass limits for SN progenitors, prior to SN 2003gd, with pre-explosion imaging.  The interpretation of these results is complicated by the different progenitor scenarios which give rise to each CCSN type.\\
The progenitor mass limits for the type II SNe (1999an, 1999br and 1999ev) are consistent with the previously determined mass limits for the type II-P SNe 1999em, 1999gi and 2001du \citep{smartt01du}.  It has been assumed that the progenitors of these SNe arise from red supergiant progenitors.   Only in the case of SN 1999br, however, has a sub-type been firmly established by long term photometric and spectroscopic monitoring \citep{past99br}.  Reported photometry from amateur observers suggests that 1999an and 1999ev were also type IIP SNe.  
The upper mass limit for the progenitor of SN 1999an does not place a tight constraint on the mass range of progenitors that give rise to type II-P SNe.  Previous estimates of the mass of the progenitors of these SNe \citep{smartt01du, smartt03gd} place the mass in the range $8 \le M_{ZAMS} \le 15M_{\odot}$.  The observed mass for the progenitor of SN 1999ev shows that type II-P SNe also arise from higher mass progenitors.  This would be inconsistent with the models of \citet{zamp99br}, for SN 1999br, if all such high mass progenitors are predicted to only give rise to faint SNe.\\  
It has not been possible to place constraining limits on the progenitors of the type I SNe (2000ds, 2000ew and 2001B) from their pre-explosion imaging.  For SNe 2000ew and 2001B the observations do not explicitly exclude a red progenitor.  The pre-explosion imaging for 2000ds, in both the V and I bands, allows the possibility of a red progenitor to be considered and ultimately excluded.  In a single star star scenario the progenitor would be star with $M_{ZAMS}>25M_{\odot}$ in the WR phase \citep{izzgrb}. The available pre-explosion F555W/F606W imaging is biassed against the discovery of the hot, low luminosity WR stars, which would require long U and B band imaging to detect.   In addition \citet{izzgrb} describe the evolution of a low mass progenitor ($10-20M_{\odot}$) in binaries, stripped by a companion, in the form of a naked helium star.  These would have temperature $\sim100,000K$ and would thus appear on the blue side of the HR diagram and would be similarly undetectable.  The location of SN 2000ds, in an S0 galaxy or earlier, strongly suggests that this SN arose in a binary system.  The pre-explosion imaging does not permit the discrimination between SNe arising from the high or low mass progenitor scenarios.
The pre-explosion imaging does, however, permit the exclusion of stars on the high mass end of the red/yellow supergiant branch at $M_{ZAMS}\approx25M_{\odot}$.\\ 
  The nature of the progenitor just prior to explosion depends on processes such as convection, mass loss, rotation and metallicity at various stages of the stars' lives.   The way these processes are implemented in the stellar evolution codes affects the predicted properties of the progenitors.
\citet{eld04} notes the inclusion of convective overshooting reduces the initial mass for the progenitor of SN 2003gd from $10$ to $8M_{\odot}$.  The utilisation of different mass loss prescriptions also causes differences between the predictions of different stellar evolution codes.  The non-rotating Geneva stellar evolution code, used here, predicts that increasingly more massive stars (up to $25M_{\odot}$) should give rise to hotter, more luminous progenitors culminating in a yellow supergiant (YSG) progenitor.  More massive stars will undergo a WR phase and will explode as type Ibc SNe.  \citet{eld04} find that, in the mass range $8-25M_{\odot}$, more massive red supergiant progenitors are increasingly more luminous, but slightly cooler.  This is consistent with observations of LMC supergiants by \citet{masslmc03}.
The use of different mass loss prescriptions causes the IIP$\rightarrow$IIL$\rightarrow$IIb$\rightarrow$Ibc transition zone to move (see figure \ref{snpops}; \citealt{heg03}; \citealt{eld04}).  \citet{hirschi04} shows that the inclusion of rotation can lead to YSG, having undergone a blue loop, and blue supergiant (BSG) progenitors for stars with $M_{ZAMS}\approx20M_{\odot}$ with rotational velocities of $\mathrm{300kms^{-1}}$.  Pre-explosion observations with the F555W and F606W filters with WFPC2 are not sensitive to BSGs, but do exclude YSG progenitors for this sample of SNe.  Direct detections of the progenitors of CCSNe will allow the different approaches of different stellar evolution models to be tested.\\
Purpose acquired post-explosion imaging of these SNe has permitted, in this study, the confident identifications of some of the SNe on pre-explosion frames.  In the specific cases of SNe 1999br, 1999ev, 2000ew and 2001B this has enabled the correct identification of the progenitor position in crowded regions on the pre-explosion imaging and the exclusion of nearby stars as candidates for the progenitor \citep{vandykprog}.  In two cases, however, the SN was not recovered on the post-explosion imaging due to a large intervening period between the SN explosion and the ACS/WFC observations.  Target of Opportunity override programs with HST will be utilised to acquire the necessary post-explosion imaging for future nearby CCSNe to provide confident identification of progenitors.
The analysis of the progenitors of the objects studied here was limited by the depth and spectral coverage of the pre-explosion data available (generally short single filter images).  Programs to acquire deep multi-wavelength imaging of nearby galaxies are currently underway, to improve the chances of detecting the progenitors and analysing their parameters.
\section{Conclusions}
We have presented mass estimates for six progenitors of CCSNe, from 
pre-explosion HST WFPC2 observations.  The identification of the progenitors 
on the pre-explosion imaging was aided by high resolution post-explosion 
imaging with the HST ACS/WFC.  SNe 1999br, 1999ev, 2000ew and 2001B were 
recovered in the HST post-explosion imaging.  The progenitor for SN 1999ev 
was confidently identified on F555W pre-explosion imaging, consistent with 
the progenitor expected for a star with $M_{ZAMS}=15-18M_{\odot}$.  The 
progenitors for 1999an, 1999br, 2000ds, 2000ew and 2001B were below the 
$3\sigma$ detection thresholds of their pre-explosion imaging.
The mass limits determined for the three IIP SNe (1999an, 1999br and 1999ev) are consistent with mass limits previously placed on SNe of this type and SN 2003gd, for which a progenitor was identified.  The observed progenitor of SN 1999ev shows that high mass stars ($>15M_{\odot}$) can give rise to normal type II-P SNe.  Our study of SN 1999br has shown that this faint SN arose from a low mass progenitor, in the same initial mass range as the progenitor of SN 2003gd, rather than a high mass progenitor.  This does not exclude the possibility of high mass progenitors, such as that of SN 1999ev, giving rise to peculiarly faint SNe.\\
The mass limits for the progenitors of the type I CCSNe (2000ds, 2000ew and 2001B) do not disagree with the predictions that these should arise from very high temperature WR stars.  In this study the mass limits determined have assumed a single progenitor scenario.  In the case of SN 2000ds a binary progenitor is much more likely, being consistent with both the lack of detection of the progenitor in the red and the age of the stellar population of NGC 2768.
The detection of a blue WR progenitor, in multiple pass bands, will be an important test of stellar evolution predictions and will be significant for the understanding of such phenomenon as GRBs.
The results presented here do, in general, concur with those of \citet{vandykprog}.  We argue, however, that post-explosion imaging of SNe 1999br, 1999ev 2000ew and 2001B allows for the confident exclusion of nearby objects as the progenitors of these SNe.  This demonstrates the importance of high resolution follow-up post-explosion imaging for this type of study.
\section*{Acknowledgments}
JRM acknowledges financial support, in the form of a studentship, from PPARC.  
This paper makes use of data obtained from the APM online sky catalogues
which are maintained by the Cambridge Astronomical Survey Unit at the
Institute of Astronomy, Cambridge.
Based on observations made with the NASA/ESA Hubble Space Telescope, obtained from the data archive at the Space Telescope Science Institute. STScI is operated by the Association of Universities for Research in Astronomy, Inc. under NASA contract NAS 5-26555.
This research has made use of the NASA/IPAC Extragalactic Database (NED) which is operated by the Jet Propulsion Laboratory, California Institute of Technology, under contract with the National Aeronautics and Space Administration.
JRM and SJS thank A. Pastorello, for the provision of TNG data, W.-R. Hamann, for the provision of the Potsdam Wolf-Rayet Models, and A.Mackey, J. Eldridge and N. Trentham for useful discussion. 

\bibliographystyle{mn2e}

\end{document}